# You Only Search Once: A Fast Automation Framework for Single-Stage DNN/Accelerator Co-design

Weiwei Chen[1,2], Ying Wang[1], Shuang Yang[1,2], Chen Liu[1], Lei Zhang[1]
[1]Institute of Computer Technology, Chinese Academy of Sciences, Beijing, China
[2]University of Chinese Academy of Sciences, Beijing, China
Email: {chenweiwei, wangying2009, yangshuang2019, liucheng, zlei}@ict.ac.cn

*Abstract*—DNN/Accelerator co-design has shown great potential in improving QoR and performance. Typical approaches separate the design flow into two-stage: (1) designing an application-specific DNN model with the highest accuracy; (2) building an accelerator considering the DNN specific characteristics. Though significant efforts have been dedicated to the improvement of DNN accuracy; it may fail in promising the highest composite score which combines the goals of accuracy and other hardware-related constraints(e.g., latency, energy efficiency) when building a specific neural network-based system. In this work, we present a single-stage automated framework, YOSO, aiming to generate the optimal solution of software-and-hardware that flexibly balances between the goal of accuracy, power, and QoS. YOSO jointly searches in the combined DNN and accelerator design spaces, which achieves a better composite score when facing a multi-objective design goal. As the search space is vast and it is costly to directly evaluate the accuracy and performance of the DNN and hardware architecture in design space search, we propose a cost-effective method to measure the accuracy and performance of solutions under consideration quickly. Compared with the two-stage method on the baseline systolic array accelerator and state-of-the-art dataset, we achieve 1.42x~2.29x energy reduction or 1.79x~3.07x latency reduction at the same level of precision, for different user-specified energy and latency optimization constraints, respectively, and the whole search procedure can be finished within 12 hours on single-card GPU.

## I. INTRODUCTION

As Deep Neural Networks (DNNs) are entering into the area of mobile and IoT for applications like visual and audio data analysis, real-time and energy-efficient hardware solutions for neural networks inference are becoming urgent needs for edge data processing [1,19,20]. Researchers in both the algorithmic and architecture communities are conducting an intensive study to create such a system for edge AI. However, there exists a wide gap between the neural network(NN) architecture design and the hardware architecture design. DNN model designers put more emphasis on improving the accuracy of a specific application and sometimes ignore the computational overhead of DNN on real devices. Even when recently researchers on DNNs are proposing lightweight DNN architectures for mobile and low power usage [8,14,17], they mostly optimize the indirect performance indicator like parameters size and operation numbers of the DNN, which will not necessarily lead to optimal performance on specific hardware. On the other hand, DNN accelerator designs [10,18] sometimes are not fully aware of the network architectures running on them, For example, early accelerator designs focus on accelerating NN models such as VGG16 or AlexNet, and they failed to provide the best hardware utility for state-of-the-art network models like MobileNetV2 or Xception. Therefore, in order to achieve the optimal performance and energy-utility in systems that are designed to run a specific or a specific domain of AI applications, DNN/accelerator co-design coordinating the efforts of algorithm and accelerator is necessary.

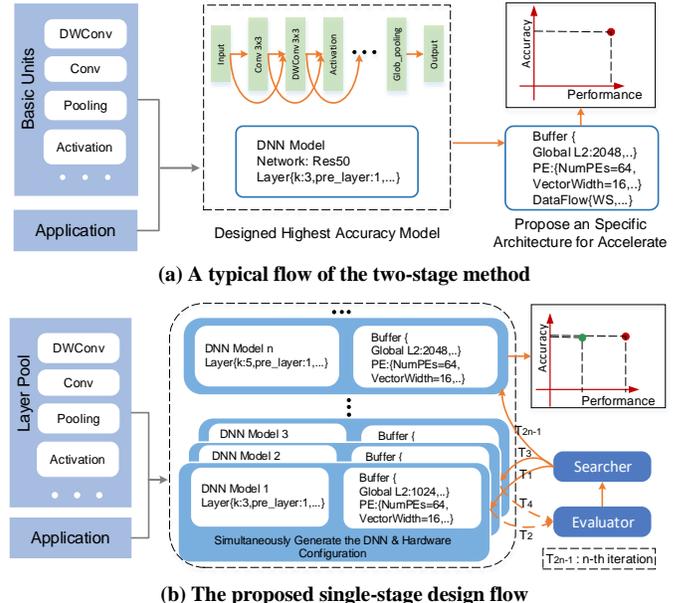

(a) A typical flow of the two-stage method

(b) The proposed single-stage design flow
Figure 1. Illustration of two-stage and our single-stage design flow.

We found recent approaches [1-4,11,12] of DNN/ accelerator co-design commonly fall into two categories. The first one is merely designing networks either manually or automatically targets to specific hardware [11,12], and the second is to follow a more complicated two-stage method [1-4], as shown in Fig 1(a). First, the two-stage co-design flow will evaluate and choose some efficient basic operation units or blocks, which are selected and stacked to construct a DNN architecture with the highest accuracy for the target dataset. Then it will customize the hardware parameters for the chosen blocks and finally form the best accelerator architecture for the model in terms of performance or energy-efficiency [1-4]. Though the two-stage solution optimizes the accelerator architecture for the highest-accuracy DNN model, it cannot guarantee the optimal software/ hardware design in cases when the design goal is more complicated than the single goal of model accuracy. For example, some mobile systems or real-time computing systems are facing more than one constraints such as latency, power, and accuracy, or need to be gauged by a composite score which combines the accuracy and other constraints[1,11,14]. In addition, in two-stage methods, there is no feedback from hardware performance indicator to the design of DNN architectures, because the two-stage co-design flow follows an uni-direction of optimization procedure from DNN to the hardware, and perhaps fails the optimality due to the lack of bi-direction software/hardware co-ordination.

Thus, our single-stage DNN software/hardware co-design framework is to guarantee system-level optimality for systems facing more than the goal of accuracy. The merit of this framework is straightforward: we jointly search in the "2-dimensional" design space where any combination of software and hardware choices are considered so that the optimality is probably achievable. However, achieving fast and efficient single-stage DNN/accelerator co-design faces two main challenges. First, the problem of single-stage co-design is more complicated than previous works on NAS [5,6,7,15]. It faces the massive "2-dimensional" design space of DNN hyper-parameters and hardware design parameters. For such a vast space, it is intractable to directly enumerate the combination of all possible DNN and accelerator design choices; effective search strategy is needed to land the optimal co-design should be able to work for any user-defined system metrics of a target system. Second, in addition to the large design space, in space search evaluating the candidates of software/hardware design that must meet any user-defined system metrics, for example, the criteria of Quality of Result (QoR), power and performance, is very time-consuming. Previous works on NAS shed light on how to efficiently converge to the target network hyper-parameters in the solution search procedure; however, they do not address the complex issue of hardware parameter search and performance evaluation, because accelerator performance modeling is tedious before practical hardware is available.

In this work, we introduce an automatic single-stage DNN/accelerator co-design framework, YOSO, to search in the vast design space at high speed. As shown in Fig 1(b), we use a reinforcement learning (RL)-based searcher to automatically search in the high dimensional space. In iterative search, the controller directly generates a combination of a network model and a hardware configuration in cooperation, which is different from all prior works and why we derived the name of *single-stage* co-design, then it promptly gets a reward from the solution evaluator and approaches the direction of bigger reward in the next iteration. The contributions are summarized as follows:

- We propose an RL+LSTM based automation framework, YOSO, which directly search in the combined DNN architecture and accelerator design space to obtain a better result than the typical two-stage method. This framework is easily transferable to different applications and supports more complicated multi-objective design optimization than prior works.

- We propose a method to quickly evaluate the neural network and performance with minimal evaluation cost. Notably, (1) we design an auxiliary HyperNet that directly generates the weights of a DNN, the HyperNet is trained with uniform sampling policy, and can fairly and efficiently evaluate the DNN's accuracy at the cost of single testing run; (2) we adopt the Gaussian Process model as the hardware performance predictor to replace the original time consuming simulation. The proposed methods help us achieve efficient search within 12 hours on single-card GPU.

- We compare our single-stage co-design method with the typical two-stage method on the baseline systolic array accelerator and Cifar10 dataset, we achieve 1.42x~2.29x energy reduction or 1.79x~3.07x latency reduction at the same level of precision, for different user-specified energy and latency optimization constraints, respectively.

## II. RELATED WORKS

As the size and depth of neural networks grow exponentially, lots of network compression techniques and different light-weight network architectures are proposed to reduce the computational complexity on IoT systems. To avoid the reliance on human efforts and experience, Neural Architecture Search (NAS) is also proposed to automatically search for efficient network architectures, such as [5,6,7,15], which is particularly imported for resource-constrained mobile [11] or embedded devices [12]. In addition to efficient network architectures, forging customized hardware accelerators to accelerate DNNs is also a popular approach to improve the AI system efficiency. Among these works, systolic array architectures (e.g. [10] [21]) are popular options for the scalability and versatility of supporting different types of dataflows. They use an array of relatively simple processing elements (PEs) to achieve high computational parallelism. However, how to select a suitable systolic array configuration and dataflow for a domain-specific system running a single or a limited set of neural network models still needs a lot of expertise and experimentation. In this work, without loss of generality, we use popular the systolic array architecture as the basic hardware accelerator template to demonstrate the proposed co-design methodology.

To pursue complicated system design goals that need to strike a balance between QoR and performance, some recent works resort to neural network and accelerator architecture co-design approach. Kiseok et al. [2] present a manual co-design of the DNN and accelerator for an embedded vision task. They firstly design a special NN accelerator intended to accelerate the SqueezeNet and then adjust the hyper-parameters of SqueezeNet to make it more efficient on the accelerator. Yifan et al. [3] adopt an algorithm-hardware co-design approach to accelerate a ConvNet model on the embedded FPGA. They employ basic 1x1 convolution blocks to form the model and design highly customized computing units based on FPGA, to boost the inference speed. However, the method is also ad-hoc for the presented system and not portable to different applications and platforms. Yu et al. [20] propose the toolchain Neutrams that transforms an existing NN to satisfy the hardware constraints of a neuromorphic chip, then finalize the hardware configuration and map the modified NN model to the neuromorphic chip. Hao et al. [1] move a step forward, and propose an FPGA/DNN co-design methodology as a two-stage approach: a hardware-oriented neural block organizer for the purpose of high-accuracy network search, and a top-down FPGA accelerator design that customizes specialized IP instances for each different neural blocks (e.g., Conv, Pooling) and then weave the IP instances together to run the organized model. These works demonstrate the performance potential of DNN/accelerator co-design. However, they mostly follow the two-stage design flow: first, they either modify an existing DNN model or design a hardware-oriented DNN model with manual or NAS for high accuracy, then design a specific accelerator that works best for the selected DNN model. The two-stage design flow limits their co-design method to search design points in a local space, which may not converge to the optimal solution, especially when we need to consider a complex design goal concerning more factors (e.g., bandwidth, QoS)[4,5,14] that creates a broader space. Besides, the design flow of hardware is for a specific application and accelerator architecture, for example, the manually customized IPs for neural blocks as in [1], we still need lots of engineering maneuver if transferred to other application or dataset.

## III. SINGLE-STAGE DNN/ACCELERATOR CO-DESIGN FLOW

### A. Problem Definition

Give the DNN architectures search space $\eta=\{\eta^1,\cdots,\eta^m\}$ that includes $m$ candidate DNN architectures; the accelerator architecture configurations search space $C=\{c^1\cdots,c^n\}$ that includes $n$ configurations; the user-provided performance constraints threshold *thres*. Our object is to automatically generate the DNN architecture $\eta*$ with associated accelerator configuration $c*$ that satisfies the *thres* while achieving the maximum accuracy $A$ for a machine learning task on the given dataset. *Perf* is the hardware performance function.

$$(\eta^*, c^*) \in \underset{\eta^* \in \eta, \; c^* \in C}{\arg\max} A(\eta, c) \quad (1)$$
$$\text{s.t.} \; Perf(\eta^*, c^*) < thres$$

The key variables in the formula, i.e. the configurable parameters of the accelerator and the basic blocks used to build the neural network, are shown in Table 1. The details of these variables are explained in the next chapters.

### B. A High-level Overview of the Automated Framework

We give a high-level overview of our single-stage framework in Fig 2. As mentioned before, our automation framework mainly solving two challenges: (1) *Tremendous search space*. (2) *Costly solution evaluation*.

For the former issue, we develop an LSTM-based RL searcher, which efficiently search in the most rewarding direction in the design space. Compared to typically search methods such as Bayesian Optimization, Bandit algorithms that behave like random search in high dimensional search space [15], the search efficiency of the adopted searcher is significantly boosted by avoiding irrelevant candidates. For the latter, we propose an effective method to evaluate the QoR and hardware performance quickly. For QoR evaluation, prior works mandate the full-training of the candidate DNN architectures in order to evaluate their accuracy on test dataset. However, we build an auxiliary HyperNet that directly generates the weights of DNN candidate to bypass the expensive procedure of fully-training. For hardware performance evaluation, we utilize machine learning techniques to create performance models without performing long-time hardware simulation. The proposed approaches allow us to identify approximately $10^6$ highly relevant hardware/software implementations from $10^{15}$ possible solutions in few hours, so that the problem of complete algorithm-hardware design space search is resolvable. The YOSO is carried out in three steps:

*Step 1: Fast evaluator construction.* The first step takes the target machine learning task (e.g., classification, detection), the basic accelerator architecture (e.g., systolic array architecture) and the user constraints (the accuracy, power and latency threshold) as inputs, and then train a HyperNet used to derive different network architectures in the search stage. After that, performance samples are taken from the accelerator simulator and used to build a performance predictor. In this work, we use energy and latency as the performance metrics for demonstration.

*Step 2: Effective design search.* An LSTM based RL searcher keeps generating the solutions iteratively, which includes the NN architecture and hardware configuration, then it receives the QoR and performance results from the evaluator to obtain the multi-objective reward, and finally update the controller towards the most rewarding design search direction.

*Step 3: Determining the final solution.* After the search process reaches a certain number of iterations, we accurately eva-luate the top-N promising candidates with the hardware simulation and fully-training, and select the best one as the final solution as output.

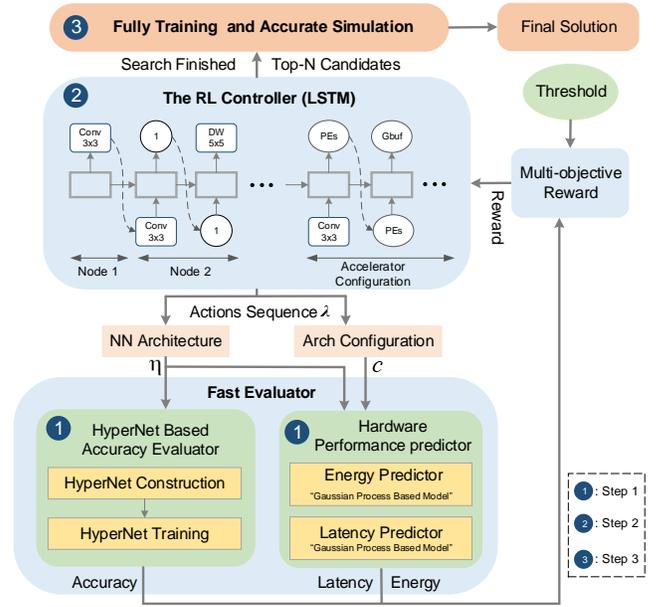

**Figure 2.** The high-level overview of our single-stage framework, YOSO.

### C. Reinforcement Learning Based Search Strategy

The RL searcher network is an LSTM with 120 hidden units, as we treat this combined optimization problem as a sequence generation task. Each candidate solution in the search space can be treated as a sequence, which is a concatenation of the DNN architecture (hyper-parameters) and the accelerator configurations, denoted as: $\lambda=(\eta,c)=(d_1,\ldots d_s, d_{s+1},\ldots d_{s+L}) \in \Lambda^{S+L}$, where $S$, $L$ are the number of hyper-parameters to present a DNN architecture and an accelerator configuration, respectively. In our experiment, $\lambda$ consists of 44 hyper-parameters (where $S$=40, $L$=4). Each parameter in the $\lambda$ can be treated as an action, so the whole sequence $\lambda$ is the actions sequence generated by the RL agent in the long term run. The LSTM samples actions via a softmax classifier in an autoregressive flow: when generating the *i-th* parameter of $\lambda$, previously generated parameters are fed as input. At the initial step, we feed zero as input. The RL can capture the relationship from the reward that can reduce the search attempts. The search cost is reduced by avoiding search solutions that have apparently inferior performance. To guide the search direction, we use a multi-objective reward signal $R(\lambda)$.

$$R(\lambda) = A(\lambda) + \alpha_1 [l(\lambda)/t\_lat]^{\omega_1} + \alpha_2 [e(\lambda)/t\_eer]^{\omega_2} \quad (2)$$

where $A(\lambda)$, $l(\lambda)$, $e(\lambda)$ are the three metrics referring to the *accuracy on the validation set*, *latency*, and *energy consumption* of $\lambda$, $t\_lat$, $t\_eer$ are the latency threshold, energy threshold, $\alpha_1$, $\omega_1$, $\alpha_2$, $\omega_2$ are four application-specific constants. Consequently, the goal is to find the hyper-parameter $\lambda^*$ that maximize the expected cumulative reward R it receives in the long run.

$$\lambda^* = \underset{\lambda \in \Lambda}{\arg\max}(E_{p(\lambda;\varphi)}[R(\lambda)]) \quad (3)$$

$\varphi$ presents the parameters of LSTM that need to be learned in the search process, $p$ is the probability to select the $\lambda$. The REINFORCE algorithm is adopted to update the LSTM para-meters, with a moving average baseline to reduce the variance.

$$\nabla_\varphi L(\varphi) = -E[(R(\lambda)-b)\nabla_\varphi \log p_{(\lambda;\varphi)}(\lambda)] \qquad (4)$$

$L$ is the loss function, $b$ presents the average baseline. It is very effective to insert the average baseline mechanism that reduces the variance of gradient estimation while keeping the bias unchanged [7], which can significantly expedite the search.

### D. HyperNet Based Accuracy Evaluator

For fast DNN model accuracy evaluation, this work builds a HyperNet, in which every NN candidate is a single path of the HyperNet and inherits the weights from the HyperNet. Therefore, the evaluation follows a typical one-time cost flow: we need to train the auxiliary HyperNet first and for once only, then obtain the accuracy of each NN candidate at the cost of a single test run.

**HyperNet Structure.** Figure 3 gives a glimpse of the HyperNet on image classification. Inspired by the works of [5,7,11,15] that construct the NN architecture by blocks; the HyperNet is stacked by two kinds of blocks: normal cell and reduction cell. They have a similar structure but with different feature shapes (each function in reduction cell has stride of 2 while the norm cell is 1). Every cell receives input from the previous two cells. When zooming into the cell, it is a directed acyclic graph consisting of an ordered sequence of B nodes(in this work, we use 7 nodes). Each node presents a latent representation (e.g., the feature map) and have the same feature shape in one cell. Each edge is associated with an operation from the candidate operations set. Each node is computed based on two previous feature nodes:

$$I_i = \theta_{(i,j)}(I_j) + \theta_{(i,k)}(I_k) \quad s.t.\ j<i\ \&\ k<i \qquad (5)$$

where $I_i$, $I_j$, $I_k$ is the $i$-th, $j$-th, and $k$-th nodes, respectively. $\theta_{(i,j)}$ and $\theta_{(i,k)}$ indicate two operations between nodes. $I_0$ and $I_1$ nodes are the outputs of previous two cells. The output of the cell is the concatenation of these nodes that do not give input to other nodes. In this work, we only use Relu as the activation function; 6 operations are included in the operations set: conv3x3, conv5x5, DWconv3x3, Dwconv5x5, max pooling, average pooling. So there exists $(6 \times (B-2)!)^4 \approx 5 \times 10^{11}$ candidate DNN architectures in the search space.

**HyperNet Training Strategy.** As HyperNet is memory costly if directly trained, we uniformly sample one sub-model (one path) in the HyperNet to perform training and only update the parameters of the selected paths in the HyperNet in each iteration, as shown in the right of Fig 3. Sampling one sub-model from HyperNet is to sequentially select some of the nodes in the cell from bottom to top. Thus, to uniformly sample the network path, suppose one node is to be selected, then it has to make two decisions in order to choose the network path from the input to the output: 1) choose two previous nodes as inputs and 2) apply two operations to the chosen inputs accordingly.:

$$p(I_0) = \cdots = p(I_{i-1}) = 2/i\ \&\ p(\theta_{(i,j)}) = 1/6 \qquad (6)$$

$j$ is the selected node, $p(I_{i-1})$ is the probability to choose the previous $i-1$ node, $p(\theta_{(i,j)})$ is the probability to choose operation $\theta_{(i,j)}$. Our experiment shows that: applying a uniform sampling strategy to HyperNet training plays a vital role in reflecting the true accuracy relation between models. If the sampling strategy is biased, such as [5,11,15], the less frequently trained sub-models are more likely to perform worse than the frequently sampled sub-models, which confuses the HyperNet to rank the sub-models by accuracy.

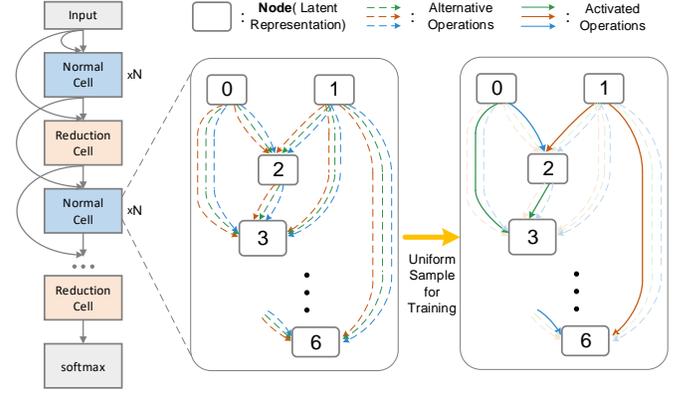

**Figure 3. A glimpse of single-shot NAS HyperNet on image classification problem and training strategy.**

### E. Gaussian Process Model-Based Cost Predictor

In classic accelerator design flow, a cycle-accurate simulator is usually built to estimate the performance before hardware implementation. The simulator models the micro-architecture with sufficient details in order to reflect the true hardware operating state, resulting in a huge time cost which is unaffordable in large design-space search. For fast cost evaluation given a network and the accelerator configuration, we adopt the machine learning technique to construct a hardware performance predictor, which achieves nearly 2000x speed improvement with less than 4% accuracy loss. In our experiment, we only build the energy and latency predictor, but this approach can be applied to other hardware performance factors.

**Energy Predictor.** As shown in Figure 4, we compare six regression models for hardware energy prediction. We collect 3600 samples from the simulation, the DNN model and configuration parameters are the input variables in these prediction models. The Gaussian Process(GP) regressor is chosen as our hardware performance predictor, for it has the lowest mean square error(MSE) among all six regression models. There are some details about the GP predictor $y$:

$$y(\lambda) = f(\lambda) + \varepsilon \qquad (7)$$

where $f(\cdot) \sim GP(\mu(\cdot), K(\cdot, \cdot))$ is the posterior distribution and $\varepsilon \sim N(\cdot \mid 0, \sigma^2)$ is Gaussian noise. $f(\cdot)$ is drawn from a Gaussian process with mean function $\mu(\cdot)$ and covariance function $K$. We use a Radial Basis Function (RBF) kernel for $K$:

$$K(\lambda, \lambda') = \exp(-\frac{1}{2\tau^2} \|\lambda - \lambda'\|^2) \qquad (8)$$

for some $\tau > 0$. In our experiment, RBF kernel works well for depicting the high-dimensional input correlation; the GP-based predictor offers reliable prediction and sampling efficiency.

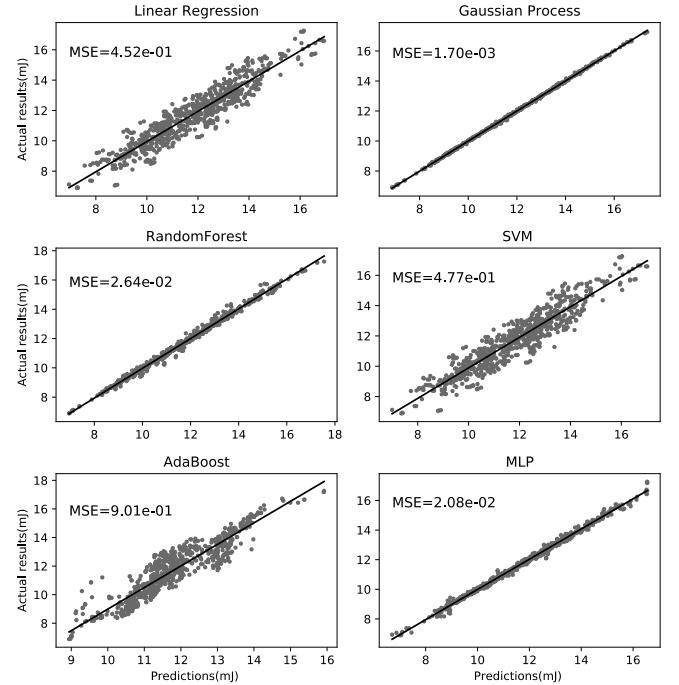

**Figure 4. Comparison between different machine learning based regression models for hardware performance predictor, for energy prediction in this example. Every model is built with 3000 training samples and tested on 600 testing samples. MSE refers to the mean squared error between prediction and actual results.**

**Latency Predictor.** Same as the energy predictor, we compare several different regression models and surprisingly find the GP predictor still works best. Therefore, we also build a GP latency predictor with different hyper-parameters.

Table 1. Key variables for DNN/accelerator co-design

| Variables | Explanation |
|---|---|
| <N_Cells, R_cells> | Number of normal cells and reduction cells to form the network. |
| B | Number of nodes in the cell |
| <$I_j$, $I_k$, $\theta_{(i,j)}$, $\theta_{(i,k)}$> | Every node in the cell. $I_j$, $I_k$ are two previous nodes to be used as inputs. $\theta_{(i,j)}$, $\theta_{(i,k)}$ present two operations to apply to the two sampled nodes. |
| Processing Element(PE) | PE array size (range:8x8~16x32) |
| g_buf | Global buffer size (range:108~1024kb) |
| r_buf | Register buffer size (range: 64~1024 byte) |
| data_flow | Four dataflows alternatives: weight stationary(WS), output stationary (OS), row stationary (RS), and no local reuse(NLR) |

## IV. EXPERIMENTAL RESULTS

### A. Experiment setup

Suppose building a customized image classification system, we evaluate our framework on Cifar10 data set and systolic array based accelerator. For demonstration, we assume that the PE array size, global buffer size, register buffer size, and the dataflow are configurable parameters before the accelerator design is finalized. The critical design variables to be searched are listed in Tab 1. Same as other machine learning works, there are 50000 and 10000 examples in the training and test sets, respectively. Because YOSO includes the software/hardware generator and performance monitor, we implement the RL controller and HyperNet based on T*ensor-Flow 1.12* with *Cuda 9.0*. For hardware performance estimation, we collect hardware performance profiles from a modified version of *nn_dataflow* [21] simulator which measures latency and energy consumption of the single solution, then employs them to train the performance predictor with *scikit-learn 0.18*. The whole process is tested on 1 Tesla P100 GPU (with 10.6 Teraflops on fp32, 32GB HBM2 at 732 GB/s) and x86 CPU (with 96 cores, Xeon 8163 at 2.5 GHz, 256 GB System memory). We set threshold requirements for energy within 9mJ and latency within1.2ms, so that the designs that fail these goals will be screened out and only the best designs will be compared. We also set two different reward functions to guide the search focus on energy optimal and latency optimal.

### B. Effectiveness of the HyperNet

We verify the effectiveness of our HyperNet based accuracy predictor: can HyperNet have the ability to rank sub-models by generating the weights to the sub-models? In other words, sub-models with inherited weights can predict true accuracy without full-training? We first set 6 blocks (4 norm cells and 2 reduction cells) to form the HyperNet and train it for 300 epochs using a batch size of 144, we adopt a stochastic gradient descent optimizer with a momentum of 0.9 with a standard random crop data argumentation. A cosine learning rate decay strategy is applied with the learning rate range between 0.05~0.0001. Moreover, we regularize the training with L2 weight decay (4x10^-5). Fig 5(a) demonstrates the training process of HyperNet. We also demonstrate the correction between the HyperNet validation accuracy with the actual validation accuracy in Fig 5(b), suggesting that HyperNet generated weights can be used to predicate the true accuracy. We randomly choose 130 sub-models from the HyperNet then evaluate these models directly on the validation set to get the search accuracy. Finally, we thoroughly train the 130 models to get real accuracy (70 epochs for each model training). We found that the accuracy of most sampled models loaded with shared weights correlates with that of stand-alone counterpart when trained fully. Models with such inherited weights can predict true accuracy; it is unclear why the shared weights can still work well in different shared models. It is critical not to overstate this claim, for we only choose several test samples that do not cover the whole space, due to the expense of running this experiment prohibits a large number of repeat trials.

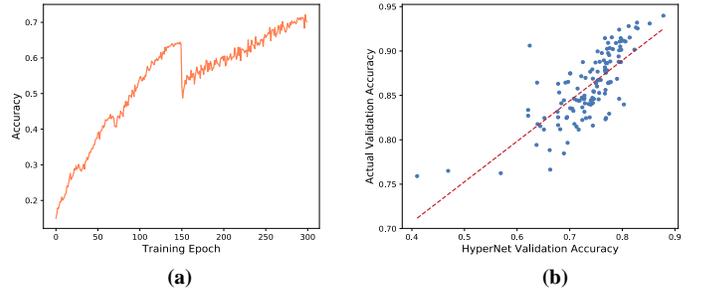

Figure 5. (a): The accuracy of HyperNet in the training process, in each epoch, we use the accuracy of a randomly sampled submodel as the accuracy of the HyperNet. (b): The correlation between the HyperNet validation accuracy with the actual validation accuracy for 130 different random NNs on Cifar10.

### C. Searching Strategy

Fig 6(a) provides the comparison of our RL based search with random search, both of the two method target on finding a higher composite score. We select every 10-th samples from 10000 iterations as examples. We use the *Reinforce* algorithm to update the RL controller. The controller is trained using Adam with a learning rate of 0.0035, to prevent premature convergence, we also use a *tanh* constant of 2.5 and a *temperature* of 1.1 for the sampling logits [7], and use the controller's sample entropy that weighted by 0.0001 to the reward. Our RL based search strategy can find better results compared with random search; it gradually finds solutions that have a higher reward score.

The coefficients in Eq.2 can be adjusted to guide the search toward different optimal regions, as preferred by different users and scenarios. For demonstration, We set two different reward functions that target to different optimal regions. We take every 20-th samples from 12000 iterations to project the search process. In Fig 6(b), we demonstrate the co-design search results are towards a user-demand tradeoff between accuracy and energy consumption. It can be seen that RL based search gradually approaches the region close to the Pareto front. Fig 6(c) demonstrates our method towards trade-off between accuracy and latency. As we can see, our RL based search strategy clearly strikes better trade-off among multi-objectives.

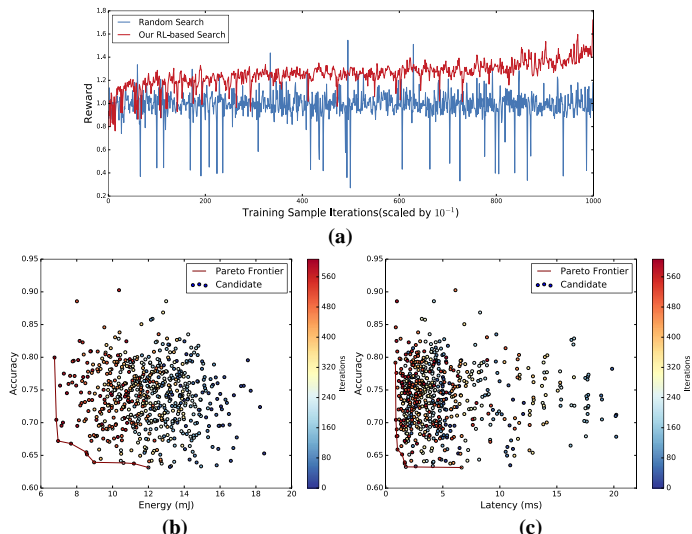

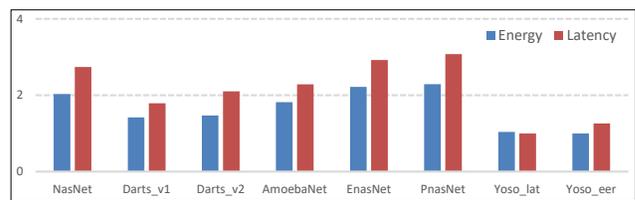

*The search time of these two-stage models we choose does not include hardware search time.

**Figure 7. Comparison of energy and latency with the two-stage method.**

**Figure 6. Demonstration of our RL based searching strategy. (a) Comparison with the random search ($α_1$:0.5 $ω_1$:-0.4 $α_2$:0.5 $ω_2$:-0.4 in Eq.2). (b): our RL method towards trade-off between accuracy and energy consumption ($α_1$:0.6 $ω_1$:-0.4 $α_2$:0.3 $ω_2$:-0.2 in Eq.2), which gradually directs its target towards the region that has higher accuracy-energy combined score. (c): our method towards trade-off between accuracy and latency ($α_1$:0.3 $ω_1$:-0.3 $α_2$:0.6 $ω_2$:-0.4 in Eq.2). All tested with threshold requirements: $t\_eer$: 9mJ $t\_lat$:1.2ms, Average search runtime is near 12 hours on one P100 GPU.**

## V. CONCLUSION

We present a fast automation framework that directly searches in the combined DNN/accelerator co-design space. As the search space is tremendous, and it is very costly to evaluate every candidate in the search space. We use RL to treat the search as a sequence generation problem to guide the search towards the Pareto frontier region. We propose the HyperNet to directly generates the weights of neural networks to avoid fully training, and a Gaussian process based predictor to predicate the hardware performance of a selected candidate. These approaches significantly improve search efficiency with fast performance evaluation, and the whole search is finished within 12 hours on single-card GPU. Compared with the two-stage method for image classification using Cifar10 targeting on systolic array based accelerator, we have achieved 1.42x~2.29x energy reduce or 1.79x~3.07x latency reduce at the same level of precision, for energy and latency optimization constraints, respectively.

### D. Comparison with Two-stage Method

We compare our single-stage search framework with the typical two-stage method. For fair comparison, We reimplement the two-stage method by choosing some existing representative neural networks that have high accuracy [5,6,9,11,13]. These networks are designed in the same neural architectures search space with ours, all the possible accelerator configuration are enumerated to select the best configuration for each network. For our single-stage method, we finish the search process with 5x10^6 iterations, then choose top-10 promising candidates to guarantee the best solution as possible, for there still exists possibility of bias in our accuracy and performance predictor. Therefore, in YOSO, the top 10 solutions will be selected, fully trained and simulated to obtain the accurate performance score for the selection of the final optimal solution. Finally, pick the best one as the delegate. Tab 2 shows the final results. *Yoso_lat* is the best solution that achieves good trade-off between accuracy and latency, which achieves the minimum latency of 0.77ms among all solutions. *Yoso_eer* presents the best solution that balances the trade-off between accuracy and energy consumption, which achieves the lowest energy consumption of 7.5mJ with comparable accuracy. Fig.7 shows the results, normalized to the lowest energy and latency. YOSO that searches in the combined search space have achieved 1.42x~2.29x energy reduce or 1.79x~3.07x latency reduce at the same level of precision, under the optimization constraint of energy and latency, respectively. Overall, compared to the two-stage method, YOSO is able to deliver better DNN models with more effective accelerator configuration.

**Table 2. Performance Comparison**

| Model | Search Time (GPU*Day) | Test Error | Energy cost(mJ) | Latency (ms) | Configuration (PEs/g_buf/r_buf/data_flow) |
|---|---|---|---|---|---|
| NasNet-A[6] | 1800 | 3.41 | 15.24 | 2.11 | 16*32/196KB/256b/OS |
| Darts_v1[5] | 0.38 | 3.0 | 10.63 | 1.38 | 16*32/512Kb/512b/OS |
| Darts_v2[5] | 1 | **2.82** | 11.01 | 1.62 | 14*16/256Kb/128b/OS |
| AmoebaNet-A[9] | 3150 | 3.12 | 13.67 | 1.76 | 16*32/108Kb/1024b/OS |
| EnasNet[11] | 1 | 2.89 | 16.65 | 2.25 | 16*32/196Kb/128b/OS |
| PnasNet[13] | 150 | 3.63 | 17.17 | 2.37 | 16*20/512Kb/256b/OS |
| Yoso_lat | 0.5 | 3.18 | 8.16 | **0.77** | 16*32/512Kb/512b/OS |
| Yoso_eer | 0.5 | 3.05 | **7.50** | 0.97 | 16*32/512kb/128b/OS |